\documentclass[12pt,twoside]{article}
\usepackage[mathscr]{eucal}
\usepackage{amsmath,amsfonts,amssymb,amsthm}
\usepackage[matrix,arrow]{xy}
\usepackage{mathabx}
\usepackage{amsthm,amsmath,latexsym,amssymb,amsfonts,amsbsy}
\usepackage{wrapfig}
\usepackage{graphicx}
\usepackage{float}

\usepackage[breaklinks]{hyperref}
\usepackage{amsmath}
\usepackage{amssymb}
\usepackage{braket}
\usepackage[T1]{fontenc}

\usepackage{tikz,pgf}
\usetikzlibrary{shapes}
\usetikzlibrary{calc}
\usetikzlibrary{decorations.pathmorphing}
\usetikzlibrary{decorations.pathreplacing,shapes.misc}
\usetikzlibrary{positioning}
\usetikzlibrary{arrows}
\usetikzlibrary{decorations.markings}

\def\be{\begin{equation}}
\def\ee{\end{equation}}
\def\ba{\begin{array}}
\def\ea{\end{array}}

\def\dps{\displaystyle}
\newcommand{\half}{\frac{1}{2}}

\def\1{\tilde{1}}
\def\2{\tilde{2}}
\def\3{\tilde{3}}

\newdimen\tableauside\tableauside=1.0ex
\newdimen\tableaurule\tableaurule=0.4pt
\newdimen\tableaustep
\def\phantomhrule#1{\hbox{\vbox to0pt{\hrule height\tableaurule
width#1\vss}}}
\def\phantomvrule#1{\vbox{\hbox to0pt{\vrule width\tableaurule
height#1\hss}}}
\def\sqr{\vbox{%
  \phantomhrule\tableaustep

\hbox{\phantomvrule\tableaustep\kern\tableaustep\phantomvrule\tableaustep}%
  \hbox{\vbox{\phantomhrule\tableauside}\kern-\tableaurule}}}
\def\squares#1{\hbox{\count0=#1\noindent\loop\sqr
  \advance\count0 by-1 \ifnum\count0>0\repeat}}
\def\tableau#1{\vcenter{\offinterlineskip
  \tableaustep=\tableauside\advance\tableaustep by-\tableaurule
  \kern\normallineskip\hbox
    {\kern\normallineskip\vbox
      {\gettableau#1 0 }%
     \kern\normallineskip\kern\tableaurule}%
  \kern\normallineskip\kern\tableaurule}}
\def\gettableau#1 {\ifnum#1=0\let\next=\null\else
  \squares{#1}\let\next=\gettableau\fi\next}

\newcommand{\bref}[1]{\textbf{\ref{#1}}}

\def\cC{\mathcal{C}}

\def\cF{\mathcal{F}}

\def\cL{\mathcal{L}}

\def\cO{\mathcal{O}}
\def\cP{\mathcal{P}}

\def\cT{\mathcal{T}}

\def\cV{\mathcal{V}}

\numberwithin{equation}{section} \makeatletter
\@addtoreset{equation}{section}

\def\be{\begin{equation}}
\def\ee{\end{equation}}
\def\ba{\begin{array}}
\def\ea{\end{array}}
\def\d{\partial}
\def\dps{\displaystyle}

\begin{document}

\begin{flushright}
FIAN-TD-2015-15 \\
\end{flushright}

\vspace{5mm}

\begin{center}

{\Large\textbf{From global to heavy-light: 5-point  conformal  blocks}}

\vspace{3mm}

\vspace{5mm}

{\large Konstantin  Alkalaev$^{\;a,b}$ and   Vladimir Belavin$^{\;a,c}$}

\vspace{0.5cm}

\textit{$^{a}$I.E. Tamm Department of Theoretical Physics, \\P.N. Lebedev Physical
Institute,\\ Leninsky ave. 53, 119991 Moscow, Russia}

\vspace{0.5cm}

\textit{$^{b}$Moscow Institute of Physics and Technology, \\
Dolgoprudnyi, 141700 Moscow region, Russia}

\vspace{0.5cm}

\textit{$^{c}$Department of Quantum Physics, \\ 
Institute for Information Transmission Problems, \\
 Bolshoy Karetny per. 19, 127994 Moscow, Russia}

\vspace{0.5cm}

\thispagestyle{empty}


\end{center}
\begin{abstract}


We consider Virasoro conformal blocks in the large central charge limit. There are different regimes depending on the behavior of the conformal dimensions. The most simple regime is reduced to the global $sl(2, \mathbb{C})$ conformal blocks while the most complicated one is known as the classical conformal blocks. Recently, Fitzpatrick, Kaplan, and Walters showed that the two regimes are related through the intermediate stage of the so-called heavy-light semiclassical limit. We study this idea in the particular case of the 5-point conformal block. To find the 5-point global  block we use the projector technique and the Casimir operator approach. Furthermore, we discuss the relation between the global and the heavy-light limits and construct the  heavy-light block from the global block. In this way we reproduce our  previous results for the 5-point perturbative classical block obtained by means of the monodromy method.

\end{abstract}
%

\section{Introduction}

The classical conformal blocks arise in many parts of mathematical physics  such as the uniformization problem, the Fuchsian monodromy problem, the AGT correspondence with four-dimensional supersymmetric gauge theories, etc. Quite  recently, the holographic interpretation of the classical conformal blocks in the context of AdS$_3$/CFT$_2$  correspondence was found.
It has been shown that the classical conformal blocks considered in a special {\it perturbative} regime  can be described as particular geodesic graphs in the conical singularity/BTZ geometry \cite{Fitzpatrick:2014vua,Asplund:2014coa,Caputa:2014eta,Hijano:2015rla,Fitzpatrick:2015zha,Alkalaev:2015wia}. \footnote{A recent interesting development has been the study of the bulk diagram technique in the presence of the defects \cite{Hijano:2015qja,Fitzpatrick:2015foa,Ageev:2015qbz}, the semiclassical bootstrap  \cite{Chang:2015qfa}, and conformal blocks beyond the semiclassical limit \cite{Beccaria:2015shq,Fitzpatrick:2015dlt}.} The main idea behind this interpretation is that in the semiclassical limit  two fields in the correlation function with heavy conformal dimensions produce the bulk geometry  while other fields with light conformal dimensions correspond to massive point particles propagating in this background. It follows that the corresponding mechanical action is identified with the classical conformal block function.    

The holographic interpretation of the conformal blocks is rather new result which certainly brings the AdS$_3$/CFT$_2$ correspondence  to the  new conceptual level of understanding. \footnote{The analogous interpretation can also be  achieved in $d$ dimensions by virtue of  the geodesic Witten diagrams \cite{Hijano:2015zsa}.} We recall that along with the characters of the symmetry algebra, conformal blocks represent the main kinematical ingredients of any CFT while the dynamical properties of the theory encoded in the structure constants of the operator algebra. 
Therefore, the holographic interpretation of conformal blocks should be supplemented by studying  the modular properties in the bulk gravity theory (for a discussion see, \textit{e.g.}, \cite{Maloney:2007ud}). There are, of course, many open problems of both conceptual and technical nature. For example, it is not clear
how to systematically analyze conformal blocks  with different portions of heavy and light fields, where heavy fields are relevant, roughly speaking, for producing AdS-like geometry in the bulk, while light fields give rise to the dynamical content of the gravitational theory. This question brings us to the study of  $n$-point  correlation functions and respective perturbative classical blocks with any number of fields $n$. Naively, heavy fields could be holographically described as a  particular configuration of a number of BTZ black holes. However, it seems that a multi-black hole solution in $AdS_3$ gravity \cite{Coussaert:1994if} cannot be directly considered as a static background for propagating point particles associated with external/intermediate fields of conformal blocks. From the CFT side our understanding is hampered by the absence of explicit results for classical $n$-point conformal blocks.

In this paper we elaborate on the recent idea by Fitzpatrick, Kaplan, and Walters (FKW) \cite{Fitzpatrick:2015zha}  about the connection between the classical conformal blocks considered in the special perturbative regime and the global $sl(2, \mathbb{C})$ conformal blocks. FKW showed that the two types of the conformal blocks turn out to be related through the intermediate stage of the so-called heavy-light semiclassical limit. The point is that the global block can be explicitly calculated and thus the heavy-light limit allows one to find the perturbative classical blocks. Here we discuss the application of these ideas to the computation of the 5-point classical conformal blocks as a first step towards the higher-point case. In particular, using the FKW procedure we reproduce the previously found expression for the 5-point conformal classical block  obtained within the monodromy approach \cite{Alkalaev:2015lca}. 

To compute the 5-point global block we use the projection technique and find a particular representation of the global block function  in terms of  Horn hypergeometric series of two variables. As a complementary method, we use the Casimir operator approach originally elaborated in the 4-point case in $D$ dimensions by Dolan and Osborn \cite{Dolan:2011dv}. In this paper we reformulate the approach in terms of CFT$_2$ notation and generalize beyond the 4-point case. In particular, we check that the 5-point global block function solves the system of two partial differential equations which are Casimir equations in two intermediate channels.

The paper is organized as follows. In Sec. \bref{sec:global_proj} we explicitly compute the 5-point global block using the projection technique. In Sec. \bref{sec:casimir} we formulate the Casimir equations in the intermediate channels. In Sec. \bref{sec:hl} we discuss  the classical and   heavy-light blocks.   In Sec. \bref{sec:5hlb} we consider the FKW procedure in the 5-point case. Finally, in Sec.  \bref{sec:conclusion} we give our conclusions.  Appendix \bref{sec:appA} contains technical details.

\section{Global conformal block}
\label{sec:global_proj}

The Virasoro algebra contains the maximal finite-dimensional subalgebra $sl(2, \mathbb{C}) \subset Vir$ of projective conformal transformations which will be further referred to as the global conformal algebra. Let $L_m$, $m\in \mathbb{Z}$ be the $Vir$ basis elements, then the $sl(2, \mathbb{C})$ basis elements are $L_m$ with $m = 0,\pm 1$. It is crucial that the global conformal algebra is also a natural truncation of Virasoro algebra in the infinite central charge limit. Indeed, rescaling $L_m \to L_m/c$, $|m|>1$ and taking the $c \to \infty$ we are left with the global conformal algebra only
\be
\label{LLL}
[L_{\pm 1}, L_0] = \pm L_{\pm 1}\;, \qquad [L_{1}, L_{-1}] = 2 L_0\;.
\ee 
In this section we consider correlation functions invariant under the global conformal algebra. 
A field $\phi = \phi(z)$ is  conformal if it satisfies the $sl(2, \mathbb{C}) $ highest-weight relations 
$L_1 \phi = 0$, and $L_0 \phi  =\Delta \phi$, where $\Delta$ is the conformal dimension. The same relations are assumed to be satisfied in the anti-holomorphic sector. Fields $L^m_{-1}\phi$ for $m \in \mathbb{N}_0$ form an $sl(2,\mathbb{C})$ module. 

Let us consider  $5$-point correlation function of the primary fields $\phi_i(z_i)$  with dimensions $\Delta_i$, $i = 1, ..., 5$. Using the projective invariance one can fix its holomorphic  dependence as follows \footnote{In what follows we omit anti-holomorphic dependence.}
\be
\label{5pointcorr}
\langle \phi_1(z_1) \cdots \phi_5(z_5)\rangle  = G(u, v) \prod_{i<j}^5 z_{ij}^{-m_{ij}}\;,
\ee
where function $G(u, v)$ depends on two projective invariants
\be
\label{uv}
\ba{c}
\dps
u = \frac{z_{12}z_{35}}{z_{13}z_{25}}\;,
\qquad
v = \frac{z_{12}z_{45}}{z_{14}z_{25}}\;,
\ea
\ee
and
\be
\label{m}
\ba{l}
m_{12} = \Delta_1 + \Delta_2 -\Delta_3 -\Delta_4-\Delta_5\;,
\qquad
\;\;m_{13} = 2\Delta_3\;,\qquad\;\; m_{14} = 2\Delta_4\;,
\\
\\
m_{15} = \Delta_1 - \Delta_2 -\Delta_3-\Delta_4 +\Delta_5\;,
\qquad m_{25} = -\Delta_1 + \Delta_2 +\Delta_3+\Delta_4 +\Delta_5\;,

\ea
\ee
while $m_{23} = m_{24} = m_{34} = m_{35} =m_{45}= 0$. Fixing   $z_1 = \infty$, $z_2=1$, $z_5 = 0$ we find that $ u = z_3$ and $v= z_4$. The exponents are chosen so that the 4-point function is directly obtained from \eqref{5pointcorr} by taking $\Delta_3=0$ along with $\d G/\d u =0$. Indeed, in this case the right-hand side of \eqref{5pointcorr} does not depend on  $z_3$, while the projective invariance remains intact. The same is true when going further to the 3-point function: taking $\Delta_3 = \Delta_4 =0$ along with $\d G/\d u =  \d G/ \d v  = 0$ gives the standard Polyakov expression.

\subsection{Projection technique }
\label{sec:proj}

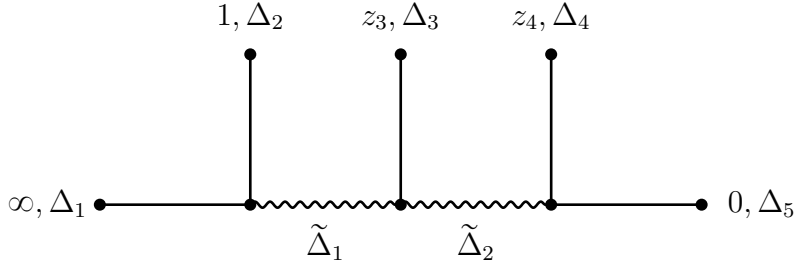
\begin{figure}[H]
\centering
\begin{tikzpicture}

\draw [line width=1pt] (30,0) -- (32,0);
\draw [line width=1pt] (32,0) -- (32,2);
\draw [smooth, tension=1.0, line width=1pt, decorate, decoration = {snake, segment length = 2mm, amplitude=0.4mm}] (32,0) -- (34,0);
\draw [line width=1pt] (34,0) -- (34,2);
\draw [smooth, tension=1.0, line width=1pt, decorate, decoration = {snake, segment length = 2mm, amplitude=0.4mm}] (34,0) -- (36,0);
\draw [line width=1pt] (36,0) -- (36,2);
\draw [line width=1pt](36,0) -- (38,0);


\draw (29.3,-0) node {$\infty, \Delta_1$};
\draw (32,2.5) node {$1, \Delta_{2}$};
\draw (34,2.5) node {$z_{3}, \Delta_{3}$};
\draw (36,2.5) node {$z_4, \Delta_{4}$};
\draw (38.8,0) node {$0, \Delta_5$};

\draw (33,-0.5) node {$\widetilde{\Delta}_1$};
\draw (35,-0.5) node {$\widetilde{\Delta}_2$};


\fill (30,0) circle (0.8mm);

\fill (32,0) circle (0.8mm);

\fill (34,0) circle (0.8mm);
\fill (32,2) circle (0.8mm);

\fill (36,0) circle (0.8mm);
\fill (34,2) circle (0.8mm);

\fill (38,0) circle (0.8mm);

\fill (36,2) circle (0.8mm);

\end{tikzpicture}
\caption{Comb diagram related to the $5$-point conformal block. The wavy lines correspond to the contributions of the two highest weight representations, 
$\widetilde \Delta_{1}$ and $\widetilde \Delta_{2}$, in the intermediate channels.}
\label{5block}
\end{figure}

To compute the global conformal block we look at the correlation function as an expectation value of the primary fields $\phi_2(z_2), \phi_3(z_3), \phi_4(z_4)$ inserted
between initial and final states $\phi_1(z_1), \phi_5(z_5)$. Using projective invariance to fix $z_1 = \infty$, $z_2 =1$, and $z_5=0$, we have
\be
\label{matrixelement}
\lim_{R \rightarrow \infty} \,R^{2\Delta_1}\,\langle \phi_1(R) \phi_2(1) \phi_3(z_3) \phi_4(z_4) \phi_5(0)\rangle = 
\langle \Delta_1| \phi_2(1) \phi_3(z_3) \phi_4(z_4) | \Delta_5\rangle\;.
\ee
A projector in the Verma module related to the primary field $\phi_{\widetilde\Delta}$ is given by
\be
\label{P}
\cP_{\widetilde\Delta} = 
\sum_{m=0}^\infty \frac{L_{-1}^m| \widetilde\Delta\rangle\, \langle \widetilde\Delta | L_1^m}{\langle \widetilde\Delta | L_1^m L_{-1}^m|\widetilde\Delta\rangle} \;,
\qquad \cP^2_{\widetilde\Delta}  = \cP_{\widetilde\Delta}\;.
\ee
Matrix element \eqref{matrixelement} with $\cP_{\widetilde{\Delta}_1}$  and $\cP_{\widetilde{\Delta}_2}$ inserted as
$G(z_3, z_4) = \langle \Delta_1| \phi_2(1)\cP_{\widetilde{\Delta}_1} \phi_3(z_3)\cP_{\widetilde{\Delta}_2} \phi_4(z_4) | \Delta_5\rangle$ defines the conformal block function
of interest 
\be
\label{formula}
G(z_3,z_4) = 
\sum_{k,m = 0}^\infty \frac{\langle \Delta_1| \phi_2(1) L_{-1}^k |\widetilde \Delta_1\rangle\, \langle \widetilde \Delta_1 | L_1^k \phi_3(z_3) L_{-1}^m | \widetilde \Delta_2 \rangle \, \langle \widetilde \Delta_2| L_1^m \phi_4(z_4) | \Delta_5\rangle}{ k! \,m!\,(2\widetilde \Delta_1)_k \,(2\widetilde \Delta_2)_m}\;.
\ee 
The corresponding  block diagram is depicted on the Fig. \bref{5block}.
All matrix elements in \eqref{formula} can be calculated explicitly (see Appendix \bref{sec:appA}). We find that 
\be
\label{series}
G(z_3,z_4) = 
 z_3^{\widetilde \Delta_1 -\Delta_3 - \widetilde \Delta_2}  \,z_4^{\widetilde \Delta_2 -\Delta_4 - \Delta_5}F(\Delta_{1,2,3,4,5}, \widetilde \Delta_{1,2}|q_1, q_2)\;,
\ee  
where new variables $q_1 = z_3$ and  $q_2 = z_4/z_3$ are introduced and 
\be
\label{coef}
F(\Delta_{1,2,3,4,5}, \widetilde \Delta_{1,2}|q_1, q_2) = \sum_{k,m = 0}^\infty F_{k,m}\, q_1^k \, q_2^m\;,
\ee
with the expansion coefficients 
\be
\label{coef2}
F_{k,m} = \frac{1}{k!m!}\frac{(\widetilde \Delta_1 + \Delta_2 -\Delta_1)_k \,(\widetilde \Delta_2 + \Delta_4 -\Delta_5)_m}{ (2\widetilde \Delta_1)_k \,(2\widetilde \Delta_2)_m}\,  \tau_{k,m}\;,
\ee
where the function $\tau_{k,m}$ is given by
\be
\label{tau2}
\tau_{k,m} = k!m!\sum_{p = 0}^{\min[k,m]} \frac{(2\widetilde \Delta_2 +m-1)^{(p)} 
(\widetilde \Delta_2+\Delta_3 - \widetilde \Delta_1)_{m-p}(\widetilde \Delta_1 + \Delta_3 -\widetilde \Delta_2+p-m)_{k-p}}{p!(k-p)!(m-p)!} \;.
\ee


First expansion coefficients (up to the second order) of the global block are
\be
\label{5ptdec}
\ba{r}
\dps
F(\Delta_{1,2,3,4,5}, \widetilde \Delta_{1,2}|q_1, q_2)=1 + \hspace{110mm}
\\
\\
\dps
+\frac{(-\Delta_1 + \Delta_2 + \widetilde{\Delta}_1) (\Delta_3 + \widetilde{\Delta}_1 - \widetilde{\Delta}_2)}{2 \widetilde{\Delta}_1}q_1  + 
\frac{(\Delta_4 - \Delta_5 + \widetilde{\Delta}_2) (\Delta_3 - \widetilde{\Delta}_1 + \widetilde{\Delta}_2)  }{2 \widetilde{\Delta}_2}q_2
 \\
\\
\dps
+\frac{(-\Delta_1 + \Delta_2 + \widetilde{\Delta}_1) (1 - \Delta_1 + \Delta_2 + \widetilde{\Delta}_1) (\Delta_3 + \widetilde{\Delta}_1 - \widetilde{\Delta}_2) (1 + \Delta_3 + \widetilde{\Delta}_1 -
    \widetilde{\Delta}_2)  }{4 \widetilde{\Delta}_1 (1 + 2 \widetilde{\Delta}_1)}q_1^2
\ea
\ee
\be
\ba{r}
\nonumber
\dps
+\frac{(\Delta_1 - \Delta_2 - \widetilde{\Delta}_1) (\Delta_4 - \Delta_5 + \widetilde{\Delta}_2) (\Delta_3 - \Delta_3^2 - \widetilde{\Delta}_1 + \widetilde{\Delta}_1^2 - \widetilde{\Delta}_2 - 
   2 \widetilde{\Delta}_1 \widetilde{\Delta}_2 + \widetilde{\Delta}_2^2)  }{4 \widetilde{\Delta}_1 \widetilde{\Delta}_2}q_1 q_2    
\\
\\
\dps 
+\frac{(\Delta_4 - \Delta_5 + \widetilde{\Delta}_2) (1 + \Delta_4 - \Delta_5 + \widetilde{\Delta}_2) (\Delta_3 - \widetilde{\Delta}_1 + \widetilde{\Delta}_2) (1 + \Delta_3 - \widetilde{\Delta}_1 + 
   \widetilde{\Delta}_2)  }{4 \widetilde{\Delta}_2 (1 + 2 \widetilde{\Delta}_2)}q_2^2+...\;.
\ea
\ee
We note that global block \eqref{coef} has the finite  Laurent part in  $z_3$. After changing to new variables $q_1$ and $q_2$  poles in $z_3$ disappear making the global block a formal power series.

The upper limit $\min[k,m]$ in \eqref{tau2} can be conveniently implemented  using  the Pochhammer symbols of negative arguments, \textit{i.e.} $(-m)_{s}$ and $(-k)_{s}$, where the summation index $s = 0,1,2,...$ . It follows that the sum in \eqref{tau2} can be  represented  in terms of the generalized hypergeometric function at 1 along with the gamma-function product, 
\be
\ba{c}
\dps
\tau_{k,m} = \frac{\Gamma (\Delta_3-\widetilde \Delta_1+\widetilde \Delta_2+m) \Gamma (\Delta_3+\widetilde \Delta_1-\widetilde \Delta_2+k-m) \, }{\Gamma (\Delta_3-\widetilde \Delta_1+\widetilde \Delta_2) \Gamma (\Delta_3+\widetilde \Delta_1-\widetilde \Delta_2-m)} \times
\\
\\
{}_3F_2(-k,-2 \widetilde \Delta_2-m+1,-m;-\Delta_3+\widetilde \Delta_1-\widetilde \Delta_2-m+1,\Delta_3+\widetilde \Delta_1-\widetilde \Delta_2-m | 1)\;.
\ea
\ee

\subsection{Horn's classification}

Using the expansion coefficients of the global block $F_{k,m}$ \eqref{coef2} we can define the ratios 
$f_{k,m} = F_{k+1,m}/F_{k,m}$ and $g_{k,m} = F_{k,m+1}/F_{k,m}$, 
which are rational functions of $k$ and $m$. The functions identically satisfy the relation $f_{k,m}g_{k+1,m} = f_{k,m+1}g_{k,m}$,  which can be taken as the definition of a hypergeometric power series in two variables  \cite{Erdel}. According to the Horn's classification, global block \eqref{coef} belongs to the trivial class of double hypergeometric series in the sense that its expansion coefficients  are represented as $F_{k,m} \sim \gamma_{k,m} R_{k,m}$, where $\gamma_{k,m}$ is a particular gamma-function product, while $R_{k,m}$ is some fixed rational function of $k$ and $m$ (see \cite{Erdel} for more details). The gamma-function product is called trivial if it is expressed either through coefficients of a single variable power series or through a product of two hypergeometric series, each in one variable. We see that the conformal block coefficients \eqref{coef2} are indeed of this form with trivial gamma-function product. 

It follows that the global block can be represented as     
\be
\label{horn}
F(\Delta_{1,2,3,4,5}, \widetilde \Delta_{1,2}|q_1, q_2) = \cT_{\Delta_3,\widetilde \Delta_{1,2}}(N_1, N_2)\Big[{}_1F_1(\widetilde \Delta_1 + \Delta_2 -\Delta_1;2\widetilde \Delta_1|q_1)\;{}_1F_1(\widetilde \Delta_2 + \Delta_4 -\Delta_5, 2\widetilde \Delta_2|q_2)\Big]\;,
\ee  
where $\cT_{\Delta_3,\widetilde \Delta_{1,2}}(N_1, N_2)$ is a rational function of  Euler operators $N_i = q_i \d/\d q_i$ giving rise to coefficients \eqref{tau2}.

Note that the  confluent  hypergeometric (Kummer's) function with parameters as in \eqref{horn} defines the OPE of two conformal fields (see, \textit{e.g.}, \cite{ZZbook}). Then the Horn's representation  \eqref{horn} is quite natural as the 5-point conformal block comb diagram shown in Fig. \bref{5block} can be cut into three pieces: two outer vertices each with two external fields and one intermediate field, and an inner vertex with one external field and two intermediate fields. Then the two confluent functions ${}_1F_1$  in \eqref{horn} correspond to the two  outer vertices, while $\cT$ corresponds to the intermediate vertex. This structure is common for any point case.

\subsection{4-point global block}
\label{sec:4ptglobal}

As already discussed, the 4-point correlation function can be obtained from \eqref{5pointcorr} simply by taking one of operators to be a unit operator, for example, $\phi_3(z_3)  = \mathbb{I}$, whence  $\Delta_3 = 0$.  In this case the fusion rules for the corresponding conformal  block say  that the intermediate conformal dimensions must be equated, $\widetilde \Delta_1 = \widetilde \Delta_2$. We find  
\be
\label{4pt}
F(\Delta_{1,2,3,4,5}, \widetilde \Delta_{1,2}|q_1, q_2)\Big|_{\footnotesize\ba{l}\Delta_3 = 0
\\ \widetilde \Delta_1 = \widetilde \Delta_2\ea} = {}_2F_1(\widetilde \Delta_1 +\Delta_2-\Delta_1, \widetilde \Delta_1 +\Delta_4-\Delta_5, 2\widetilde \Delta_1| q_1q_2)\;,
\ee
where the right-hand-side is simply the global 4--point block depending on $q_1q_2 := z_4$ \cite{Ferrara:1974ny}. Indeed, setting $\Delta_3 = 0$ and $\widetilde \Delta_1 = \widetilde \Delta_2$ we can show that $\tau_{k,m} = \delta_{km} (m!)^2 k!(2\widetilde \Delta_1)_m$, and therefore  \eqref{coef} is reduced to the hypergeometric series with parameters as in \eqref{4pt}.

\subsection{Vacuum global blocks}

If one of intermediate fields is a unit  field, then we arrive at the vacuum conformal block. In the 4-point case the vacuum block is obtained by setting the intermediate dimension $\Delta = 0$. It follows that the 4-point vacuum global   block \eqref{4pt} is trivial, 
\be
F(\Delta_{1,2,3,4},\Delta|z)\Big|_{\Delta =0} = {}_2F_1(\Delta_2-\Delta_1, \Delta_3-\Delta_4, 0| z) = 1\;.
\ee

In the 5-point case there are two vacuum blocks corresponding  either to $\widetilde \Delta_1 =0$ or to $\widetilde \Delta_2 = 0$. In what follows we find that they are given by  hypergeometric functions evaluated at particular values of parameters and reduced to power functions. 
Indeed, in this case the intermediate channel states form the trivial  $sl(2,\mathbb{C})$ module and therefore the  5-point comb diagram is disconnected. The corresponding 5-point correlator function splits into 2-point and 3-point correlators given by power functions. 

\vspace{-3mm}

\paragraph{Vacuum block I.} In this case $\widetilde \Delta_2=0$ and $\widetilde \Delta_1 = \Delta_3$, where the second condition is guaranteed by the fusion rules for the intermediate vertex.  Then, 
\be
F_{k,m} = \frac{(\widetilde \Delta_1+\Delta_2-\Delta_1)_k \,(\widetilde \Delta_2+\Delta_4-\Delta_5)_m}{ (2\widetilde \Delta_1)_k \,(2\widetilde \Delta_2)_m}\,  \tau_{k,m}\;,
\ee
where coefficients $\tau_{k,m}$ are read off from \eqref{tau2}, namely $\tau_{k,m} = \delta_{m,0}m! (2\widetilde \Delta_1)_k$. It follows that the type I  vacuum global block is 
\be
\label{tau234}
\ba{r}
F^{I}_{vac}(\Delta_1, \Delta_2,\Delta_3|q_1) = {}_2F_1(\Delta_3+\Delta_2-\Delta_1, 2\Delta_3,2\Delta_3|q_1)\equiv {}_1F_0(\Delta_1-\Delta_2-\Delta_3|q_1) 
\\
\\
=   (1-q_1)^{\Delta_1-\Delta_2-\Delta_3}\;. 
\ea
\ee

\vspace{-3mm}

\paragraph{Vacuum block II.} In this case $\widetilde \Delta_1=0$ and $\widetilde \Delta_2 = \Delta_3$, where the second condition is guaranteed by the fusion rules for the intermediate vertex. 
Then,  
\be
F_{k,m} = \frac{(\Delta_2 - \Delta_1)_k \,(\Delta_3+\Delta_4-\Delta_5)_m}{ (2\widetilde \Delta_1)_k \,(2\Delta_3)_m}\,  \tau_{k,m}\;,
\ee
and the coefficient $\tau_{k,m}$ can be read off from \eqref{tau2}. We note that in order to avoid a pole we have to set $k=0$ in the above formula. In this case the coefficient $\tau_{0,m} =(2\Delta_3)_m$, and finally we arrive at the type II vacuum global  block
\be
\label{tau234}
\ba{r}
F^{II}_{vac}(\Delta_3, \Delta_4,\Delta_5|q_2) = {}_2F_1(\Delta_3+\Delta_4-\Delta_5, 2\Delta_3,2\Delta_3|q_2) \equiv {}_1F_0(\Delta_5-\Delta_4-\Delta_3|q_2)  
\\
\\= (1-q_2)^{\Delta_5-\Delta_4-\Delta_3}\;. 
\ea
\ee

\subsection{Small $\Delta_3$ expansion} 
\label{sec:small3}

In our further analysis we are especially interested in the 5-point conformal block with particular dimensions 
\be
\label{delta3}
\widetilde \Delta_1 = \widetilde \Delta_2\;, \qquad \text{and }\qquad  \Delta_1 = \Delta_2\;, \qquad \Delta_4 = \Delta_5\;.
\ee
In Sec. \bref{sec:hl} we treat 5-point functions as a deformation of 4-point functions with respect to a small conformal dimension of the third external field \cite{Alkalaev:2015wia,Alkalaev:2015lca}. The deformation procedure is consistent provided the intermediate dimensions are set equal to each other as in the first constraint in \eqref{delta3}. Then, the intermediate vertex in the limit $\Delta_3=0$ is compatible with the fusion rules. The last two constraints in \eqref{delta3} are  for convenience only.  

The global  block with dimensions \eqref{delta3} reads 
\be
\label{222}
F(\Delta_{1,3,4}, \widetilde \Delta_{1}|q_1, q_2) = \sum_{k,m = 0}^\infty  \frac{(\widetilde \Delta_1)_k \,(\widetilde \Delta_1)_m}{ (2\widetilde \Delta_1)_k \,(2\widetilde \Delta_1)_m}  \sum_{p = 0}^{\min[k,m]} \frac{(2\widetilde \Delta_1 +m-1)^{(p)} 
(\Delta_3)_{m-p}(\Delta_3+p-m)_{k-p}}{p!(k-p)!(m-p!)}\, q_1^k \, q_2^m\;,
\ee
and the small $\Delta_3$ expansion is naturally given by 
\be
F(\Delta_{1,3,4}, \widetilde \Delta_{1}|q_1, q_2) = {}_2F_1(\widetilde \Delta_1 , \widetilde \Delta_1, 2\widetilde \Delta_1| q_1q_2) + \sum_{s=1}^\infty \Delta_3^s\, \cC_{(s)} (\Delta_{1,4}, \widetilde \Delta_{1}|q_1, q_2)\;,
\ee
where the leading term is identified with the 4-point global block \eqref{4pt}, while $\cC_{(s)}$ are sub-leading corrections that can be directly  read off from \eqref{222} as particular power series. It would be interesting to find the corrections in a closed form.\footnote{ Compare with  the 5-point heavy-light conformal block  expanded up to the forth order in classical dimension $\epsilon_3  = \Delta_3/c$ \eqref{exf}. }

\section{Global conformal block: the Casimir equations}
\label{sec:casimir}

The conformal blocks are universal functions which are completely defined by the conformal Virasoro symmetry and  apart form the central charge (which in our case goes to infinity) depend only on the conformal dimensions of the field insertions. In the case of global blocks the conformal algebra is reduced to projective $sl(2, \mathbb{C})$ subalgebra. It follows that the global conformal blocks can be  defined as eigenfunctions of the $sl(2, \mathbb{C})$  Casimir operator with eigenvalues  given by the intermediate conformal dimensions. In the 4-point case the corresponding Casimir equation reads
\be
\label{casimir}
\Big[C_2 +2 \Delta(\Delta-1)\Big]\langle \,\phi_1(z_1) \cdots \phi_4(z_4)\rangle = 0\;,
\ee
where $C_2$ is the second-order differential Casimir operator, and $\Delta$ is the intermediate conformal dimension \cite{Dolan:2011dv} (for review see also \cite{SimmonsDuffin:2012uy}). The correlation function in \eqref{casimir} should be understood as being  restricted onto the intermediate channel thereby giving the corresponding conformal block. 

In what follows we introduce the Casimir equations in the 5-point case. To this end, we recall that the global Ward identities fixing  correlation function \eqref{5pointcorr} read 
\be
\label{ward}
\Big[G^{(1)}_\alpha+ ... + G_\alpha^{(5)}\Big]\, \langle \phi_1(z_1) \cdots \phi_5(z_5)\rangle = 0\;, \qquad \alpha = 0,\pm 1\;,
\ee 
where the symmetry generators  given by $G^{(i)}_\alpha = (z^{\alpha+1}_i\d_i+ (\alpha+1) \Delta_i z^\alpha_i)$ form the $sl(2,\mathbb{C})$ algebra
\be
[G^{(i)}_{-1}, G^{(i)}_{0}] = G^{(i)}_{-1}\;, 
\quad
[G^{(i)}_{-1},G^{(i)}_{1}] = 2G^{(i)}_0\;,
\quad
[G^{(i)}_0,G^{(i)}_{1}] = G^{(i)}_1\;,
\ee
cf. \eqref{LLL}, while $[G^{(i)}_\alpha, G^{(j)}_\beta] =0$ for $i \neq j$. The $sl(2, \mathbb{C})$ Casimir operator is given by 
\be
C_2(i) = -2(G^{(i)}_{0})^2 + G^{(i)}_{-1} G^{(i)}_{1} + G^{(i)}_{1}G^{(i)}_{-1}\;.
\ee

In our case, there are  two  Casimir equations in two intermediate channels 
\be
\label{Caseq}
\Big[C_2(4,5) +2\widetilde \Delta_2(\widetilde \Delta_2-1)\Big]\langle \phi_1(z_1) \cdots \phi_5(z_5)\rangle = 0\;,
\ee  
\be
\Big[C_2(3,4,5) +2\widetilde \Delta_1(\widetilde \Delta_1-1)\Big]\langle \phi_1(z_1) \cdots \phi_5(z_5)\rangle = 0\;,
\ee  
where $\widetilde \Delta_1$ and $\widetilde \Delta_2$ are intermediate conformal dimensions. The Casimir operators $C_2(4,5)$ and $C_2(3,4,5)$ are defined as follows
\be
\label{casimirs123}
\ba{l}
C_2(4,5) = -2(G_0^{(4)}+G_0^{(5)})(G_0^{(4)}+G_0^{(5)}) + \{(G_{1}^{(4)}+G_{1}^{(5)}),(G_{-1}^{(4)}+G_{-1}^{(5)})\}\;,
\\
\\
C_2(3,4,5) = -2(G_0^{(3)}+G_0^{(4)}+G_0^{(5)})(G_0^{(3)}+G_0^{(4)}+G_0^{(5)}) + 
\\
\\
\hspace{60mm}+\{(G_{1}^{(3)}+G_{1}^{(4)}+G_{1}^{(5)}),(G_{-1}^{(3)}+G_{-1}^{(4)}+G_{-1}^{(5)})\} \;.
\ea
\ee
In fact, there are  two more equations related to the OPE of  $\phi_1$, $\phi_2$, and $\phi_1$, $\phi_2$, $\phi_3$, respectively. However, using  Ward identities \eqref{ward} their Casimir operators are equal to those in \eqref{casimirs123}, \textit{i.e.} $C_2(4,5) = C_2(1,2,3)$ and $C_2(3,4,5) = C_2(1,2)$.

\subsection{4-point global  block} 

The 4-point correlation function can be chosen in the form \eqref{5pointcorr}--\eqref{m}, 
\be
\langle \phi_1(z_1) \cdots \phi_4(z_4)\rangle= z_{13}{}^{-2 \Delta _3} z_{14}{}^{-\Delta _1+\Delta _2+\Delta _3-\Delta _4} z_{12}{}^{-\Delta _1-\Delta _2+\Delta _3+\Delta _4} z_{24}{}^{\Delta _1-\Delta _2-\Delta _3-\Delta _4}G(x)\;,
\ee
where the cross ratio is  $x = (z_{12} z_{34} )/(z_{13} z_{24})$. Fixing $z_1 = \infty, z_2 = 1, z_3 = z, z_4 = 0$, we find that the action of the Casimir operator on the 4-point correlation function is given by   
\be
\label{4ptcasimir}
\ba{l}
\dps
C_2(3,4) \langle \phi_1(z_1) \cdots \phi_4(z_4)\rangle = \lim_{z_1\to \infty}2 z_1^{-2\Delta_1}\Big(z ((z-1) z G''(z)+
\\
\\
\dps\hspace{55mm}+(-2 \Delta _3-2 \Delta _4-\Delta _1 z+\Delta _2 z+3 \Delta _3 z+\Delta _4 z+z) G'(z))+
\\
\\
\hspace{20mm}+G(z) (-(\Delta _4-1) \Delta _4+\Delta _3^2 (2 z-1)+\Delta _3 (-2 \Delta _4-2 \Delta _1 z+2 \Delta _2 z+2 \Delta _4 z+1))\Big)\;,
\ea
\ee
where $G'(z)$ and $G''(z)$ are first and second derivatives in $z$ variable. 
The Casimir equation in this case is given by \eqref{casimir}. Imposing particular boundary conditions we find that the solution is given by 
\be
\label{hyperblock}
\langle \phi_1(z_1) \cdots \phi_4(z_4)\rangle \cong  z^{\Delta -\Delta _3-\Delta _4} \, _2F_1(\Delta -\Delta _1+\Delta _2,\Delta +\Delta _3-\Delta _4,2 \Delta |z)\;,
\ee
where $\cong$ means that the correlation function is restricted to the intermediate channel with the dimension $\Delta$. Function \eqref{hyperblock} reproduces the 4-point global  block \cite{Ferrara:1974ny,Dolan:2011dv}, cf. \eqref{4pt}. It is instructive to remove the exponential prefactor in \eqref{hyperblock} by $G(v) :=  v^{\Delta -\Delta _3-\Delta _4} F(v)$ and  obtain the standard hypergeometric equation 
\be
\label{hyper}
a\frac{\d^2 F}{\d v^2}+ d \frac{\d F}{\d v} + eF = 0\;,
\ee
with coefficients  
\be
\label{hypercoef}
a = v(v-1)\;,
\quad
d = (-\Delta _1+\Delta _2+\Delta _3-\Delta _4)v+2 \Delta  (v-1)+v\;,
\quad
e = (\Delta -\Delta _1+\Delta _2) (\Delta +\Delta _3-\Delta _4)\;.
\ee

\subsection{5-point global conformal block}

The educated guess is that the Casimir equations can be conveniently represented as acting directly on the conformal block function $F(\Delta_{1,2,3,4,5}, \widetilde\Delta_{1,2}|z_3,z_4)$, cf. \eqref{hyper}. Following \eqref{series} we redefine 
\be
\langle \phi_1(z_1) \cdots \phi_5(z_5)\rangle \cong z_3^{\widetilde \Delta_1 -\Delta_3 - \widetilde \Delta_2}  \,z_4^{\widetilde \Delta_2 -\Delta_4 - \Delta_5}F(\Delta_{1,2,3,4,5}, \widetilde\Delta_{1,2}|z_3,z_4)\;,
\ee
where $\cong$ means that the correlation function is restricted to the intermediate channels with dimensions $\widetilde\Delta_{1,2}$. Denoting $u = z_3$ and $v = z_4$ \eqref{uv} we find that the Casimir equations for the conformal block function $F = F(\Delta_{1,2,3,4,5}, \widetilde\Delta_{1,2}|u,v)$  are given by two second order PDEs
\be
\label{cas1}
a \frac{\d^2 F}{\d v^2 } + b  \frac{\d^2 F}{\d u \d v } + c \frac{\d F}{\d u} + d\frac{\d F}{\d v} + e F = 0\;,
\ee
\be
\label{cas2}
m \frac{\d^2 F}{\d u^2 } + n  \frac{\d^2 F}{\d u \d v } + k \frac{\d F}{\d u} + l\frac{\d F}{\d v} + p F = 0\;,
\ee
where the coefficient are 
\be
\label{k1}
\ba{l}
a = u v (v-1)\;,\qquad b = uv(u-1)\;,
\qquad
c= u(u-1)(\Delta_4-\Delta_5+\widetilde \Delta_2)\;,
\\
\\
d = v(\Delta_3-\widetilde \Delta_1+\widetilde \Delta_2)+u v (-\Delta_1+\Delta _2+\Delta _4-\Delta_5+\widetilde \Delta_1+\widetilde \Delta_2+1)-2 u \widetilde \Delta_2\;,
\\
\\
e = (\Delta_4-\Delta_5+\widetilde \Delta_2)  (\Delta_3-\widetilde \Delta_1+\widetilde \Delta_2+u (\Delta_2-\Delta_1+\widetilde \Delta_1)) \;,
\ea
\ee
and
\be
\label{k2}
\ba{l}
m = u^3(u-1)\;,\qquad n = uv(u-1)^2+uv(v-1)\;,
\\
\\
k = u^3 (-\Delta_1+\Delta_2+\Delta_3+2 \widetilde \Delta_1-\widetilde \Delta_2+1)-2 u^2 \widetilde \Delta_1+uv(\widetilde \Delta_2
+\Delta_4-\Delta_5)\;,
\\
\\
l = u^2 v (\Delta_3+\widetilde \Delta_1-\widetilde \Delta_2)+2 uv (\widetilde \Delta_2-\widetilde \Delta_1)-v^2 (\Delta_3-\widetilde \Delta_1+\widetilde \Delta_2)\;,
\\
\\
p = -u^2 (\Delta_1-\Delta_2-\widetilde \Delta_1) (\Delta_3+\widetilde \Delta_1-\widetilde \Delta_2)-v (\Delta_4-\Delta_5+\widetilde \Delta_2) (\Delta_3-\widetilde \Delta_1+\widetilde \Delta_2)\;.

\ea
\ee
If $\Delta_3=0$ and $\widetilde \Delta_1 = \widetilde \Delta_2$ (cf. \eqref{delta3}), and  $\d F/\d u = 0$,  then the first equation reproduces  the 4-point case equation  \eqref{hyper}--\eqref{hypercoef}, while the second equation disappears.    

The 5-point global conformal block \eqref{coef}--\eqref{tau2} solves the two  Casimir  equations \eqref{cas1}--\eqref{k2}. 

\section{Heavy-light and linearized classical blocks}
\label{sec:hl}

According to the original Zamolodchikov's definition of the \textit{light} field, this is one with a fixed value of the conformal dimension $\Delta$ in the limit   $c\rightarrow \infty$. In particular, the conformal blocks for light fields are just the global blocks discussed  in the previous sections. 
However, in the context of the AdS/CFT correspondence another regime of the conformal dimensions is more relevant. It deals with the \textit{heavy} fields which are conventionally described by the \textit{classical} conformal dimensions $\epsilon =\dps\lim_{c\to \infty} \Delta/c$   fixed at $c\rightarrow \infty$. It is well-known that the presence of the heavy fields in the spectrum of the boundary ($\sim$ bulk) theory allows producing localized high energy states in the bulk of AdS$_3$ such as conical defects or BTZ black holes. All known examples show that in the semiclassical limit  $c\rightarrow \infty$ the conformal blocks  for heavy fields are exponentiated  (see, \textit{e.g.}, \cite{Zamolodchikov1986}) 
\be
\label{ccb}
\cF(\Delta_i, \widetilde \Delta_j|z_k) = \exp\big[-\frac{c}{6}f(\epsilon_i, \tilde \epsilon_j|z_k)\big]\;,
\ee
where $\epsilon_i$ and $\tilde \epsilon_j$ stand for external and intermediate classical conformal dimensions,  respectively, and
function $f(\epsilon_i, \tilde \epsilon_j|z_k)$ is called the (non-perturbative) {\it classical} conformal block. 

From the AdS/CFT perspective, it is instructive to study the  {\it linearized} version of the classical conformal block. In this regime  the  classical conformal dimensions  for some
subset of fields $\epsilon_l\ll1$ so that only lower coefficients in the $\epsilon_l$ expansion  of the classical conformal block $f(\epsilon_i, \tilde \epsilon_j|z_i)$ are relevant. 
According to \cite{Perlmutter:2015iya}, such fields sometimes are called perturbative heavy fields \footnote{Note that in some previous papers (in particular, in \cite{Alkalaev:2015wia,Alkalaev:2015lca}) these fields were referred to as light fields. Such a notation
leads to the obvious confusion with the initial Zamolodchikov's definition. However, as we discuss below, there is a non-trivial connection \cite{Fitzpatrick:2015zha} between the conformal blocks with perturbative heavy fields and light fields.} (or simply perturbative fields) and the corresponding block -- perturbative classical conformal block.

In the semiclassical limit, the dual AdS$_3$ gravity  is weakly coupled and hence allows for the saddle-point approximation described by the dual  Witten type geodesic diagrams. More precisely, the linearized  classical block is described in the bulk theory by means of classical relativistic mechanics of massive particles (corresponding to the perturbative fields) in the asymptotically AdS geometry induced by the heavy fields. 
In what follows, we are focused on the case of two heavy fields with equal classical conformal dimensions $\epsilon_h$ producing in the bulk a conical singularity with the deficit angle $\alpha=\sqrt{1-4\epsilon_h}$, while the other operators  arising in the intermediate channels are perturbative \cite{Fitzpatrick:2014vua,Asplund:2014coa,Caputa:2014eta,Hijano:2015rla,Fitzpatrick:2015zha,Alkalaev:2015wia,Hijano:2015qja}.  

Below we quote the expression of the perturbative classical conformal block found in \cite{Alkalaev:2015lca} (see also \cite{Alkalaev:2015wia}). 
Here, function $f = f(\epsilon_h, \epsilon_{3},\epsilon_4, \tilde \epsilon_1|z_3,z_4)$ denotes  the $5$-point perturbative classical  block with  $\epsilon_1 = \epsilon_2\equiv \epsilon_h$, $\epsilon_4 = \epsilon_5$ and $\tilde \epsilon_1 = \tilde\epsilon_2$ \eqref{delta3}
related by means of  \eqref{ccb}  to the quantum block depicted on the  Fig. \bref{5block}. The power series expansion of the perturbative classical block  up to fourth order in the third dimension $\epsilon_3$  is given by
\be
\label{exf}
f=f^{(0)}+\epsilon_3 f^{(1)} + \epsilon_3^2 f^{(2)}+ \epsilon_3^3 f^{(3)} + \epsilon_3^4 f^{(4)}+...\;,
\ee
where 
\begin{align}
\label{fcoef}
f^{(0)}={}&
\ln\Big[(1-a)^{2\epsilon_4}a^{(\frac{1}{\alpha}-1)\epsilon_4} \Big(\frac{\sqrt{a}+1}{\sqrt{a}-1}\Big)^{\tilde \epsilon_1}\Big]\;,\\
f^{(1)}={}&
\ln\Big[ b^{(\frac{1}{\alpha}-1)}\Big(\frac{a - b^2}{ \sqrt{a}}\Big)\Big]\;,\\
f^{(2)}={}&\Big[(a+b^2)(4 a b-a - a^2 - b^2 - a b^2)\Big]\times\big(4 a^{1/2} (a-b^2)^2\tilde{\epsilon}_1 \big)^{-1}\;,\\
\quad
f^{(3)}={}&\Big[(1- b) b (a - b) (a + b^2) (a + a^2 - 4 a b + b^2 + a b^2)\Big]\times\big(2  (a-b^2)^4\tilde{\epsilon}_1^2\big)^{-1}\;,
\end{align}
\begin{align}
f^{(4)}={}& \Big[a^6 - 9 a^7 - 9 a^8 + a^9 + 24 a^6 b + 
   144 a^7 b + 24 a^8 b - 30 a^5 b^2 - 546 a^6 b^2 -  546 a^7 b^2 - \nonumber \\&
   30 a^8 b^2 + 792 a^5 b^3 + 2576 a^6 b^3 + 792 a^7 b^3 - 
   321 a^4 b^4 - 4599 a^5 b^4 - 4599 a^6 b^4 -  321 a^7 b^4 + \nonumber  \\&
   3024 a^4 b^5 + 10080 a^5 b^5 + 3024 a^6 b^5 - 580 a^3 b^6 - 
   8892 a^4 b^6 - 8892 a^5 b^6 - 580 a^6 b^6  + \nonumber \\&   
   3024 a^3 b^7 + 10080 a^4 b^7 +  3024 a^5 b^7 - 321 a^2 b^8 - 4599 a^3 b^8 - 
   4599 a^4 b^8 - 321 a^5 b^8 +  \nonumber  \\& 792 a^2 b^9 +  2576 a^3 b^9 +
   792 a^4 b^9 - 30 a b^{10} -546 a^2 b^{10} - 546 a^3 b^{10} - 
   30 a^4 b^{10} + 24 a b^{11} + \nonumber \\&
   144 a^2 b^{11} + 24 a^3 b^{11} +  b^{12} -  9 a b^{12} - 9 a^2 b^{12} + a^3 b^{12}\Big]\times\big(192 a^{3/2} (a - b^2)^6  \tilde{\epsilon}_1^3\big)^{-1}\;,
\label{dddd}
\end{align}
and $a = (1-z_4)^\alpha$ and $b = (1-z_3)^\alpha$. Note that  expansion \eqref{exf} assumes that $\epsilon_3/\tilde\epsilon_1 \ll 1$, and thus the linear approximation is still valid. The leading term here  is the 4-point perturbative classical block \cite{Fitzpatrick:2014vua,Hijano:2015rla}. The subleading terms with non-vanishing degrees of the third dimension $\epsilon_3$ describe the  order by order deformation. 

So far we have discussed two possible classical limits of the conformal blocks: the global conformal block containing only light fields and
the (linearized or perturbative) classical conformal block containing only heavy fields. Naturally, one can consider all possible intermediate scenarios, where
the conformal block contains both light and heavy fields. This regime is known as the {\it heavy-light limit}. 
Below we denote by $\cV(\epsilon_h,\Delta, \widetilde\Delta|z_k)$ the heavy-light conformal block 
\be
\label{hlb}
\cV(\epsilon_h,\Delta_i, \widetilde\Delta_j|z_k)\coloneq\!\lim_{\substack{ c\to \infty \\ \epsilon_h,\Delta_i, \widetilde \Delta_j -\text{fixed}}} \cF (c \epsilon_h,\Delta_i, \widetilde\Delta_j|z_k)\;, 
\ee
where $\epsilon_h$ is the classical heavy dimension,  $\Delta_i$ and $\widetilde \Delta_j$ are the quantum   external and intermediate light dimensions. 

An interesting observation by Fitzpatrick, Kaplan, and Walters \cite{Fitzpatrick:2015zha} is that 
the global and classical blocks are related through the {\it heavy-light limit}. It follows that the computation of the linearized classical conformal block can be reduced  to that of the global  block.  The algorithm  consists of two steps.

\vspace{-5mm}

\paragraph*{Step 1.} We use the fact that heavy-light conformal blocks are equivalent to global conformal blocks considered in some non-trivial background metric \cite{Fitzpatrick:2015zha}. 
In the case of two heavy fields with equal classical dimensions $\epsilon_h$ this is achieved by  mapping the positions of the external light fields from $z$ to $w = z^\alpha$.  
The crucial point is that in the new coordinates in the limit $c \to \infty$ only elements $L_{-1}^k$ contribute  in the matrix elements  leaving us with the global block modulo the Jacobian prefactors~\cite{Fitzpatrick:2015zha}.

\vspace{-3mm}

\paragraph*{Step 2.} The heavy-light conformal block 
$\cV(\epsilon_h,\Delta_i,\widetilde\Delta_j|z)$ and the linearized classical conformal block $f^{\text{lin}}(\epsilon_h, \epsilon_i, \tilde \epsilon_j|z)$ are related. 
Schematically, the idea is that the following two limits can be rearranged \cite{Fitzpatrick:2015zha}. Instead of taking first the limit $c\rightarrow\infty$ with all fields  be heavy (\textit{i.e.} with fixed ratios $\Delta_i/c$) and then considering 
some subset of the fields to be small $\epsilon_i,\tilde\epsilon_j\ll1$ as we do in order to calculate  the linearized classical conformal block,
one can first take the heavy-light limit, \textit{i.e.} $c\rightarrow\infty$ with only a subset of  fields  be heavy (with classical dimensions $\epsilon_h$)
and then take  $\Delta_i,\widetilde\Delta_j\gg1$ for the light subset of the fields. Using \eqref{ccb} we find that the linearized classical block is given by 
\be
\label{lcb}
f^{\text{lin}}(\epsilon_h, \epsilon_i, \tilde \epsilon_j|z)  = \Big[\lim_{\substack{ c\to \infty \\ \epsilon_h,\epsilon_i ,\tilde\epsilon_j -\text{fixed}}} -\frac{6}{c}\log\cF(c\epsilon_h, c \epsilon_i, c\tilde \epsilon_j|z)\Big]_{\epsilon_i,\tilde\epsilon_j \ll 1}\;,
\ee
while the logarithm of the heavy-light block \eqref{hlb}  is given by  
\be
\label{loghlb}
\log \cV(\epsilon_h,\Delta_i, \widetilde\Delta_j|z) = \lim_{\substack{ c\to \infty \\ \epsilon_h,\Delta_i,\tilde\Delta_j -\text{fixed}}} \log \cF (c \epsilon_h,\Delta_i, \tilde \Delta_j|z)\;.
\ee
Using the well-known properties of the conformal blocks as rational functions of the conformal dimensions  and the central charge,  one can argue  that in the large central charge limit their  coefficients obey simple homogeneity properties giving rise  to the following relation 
\be
\label{blocks}
f^{lin}(\epsilon_h, \epsilon_i, \tilde \epsilon_j|z) =  \Big[\log \cV(\epsilon_h,\epsilon_i, \tilde \epsilon_j|z)\Big]_{\deg(\epsilon, \tilde \epsilon)=1}\;,
\ee 
where $[g(\epsilon_i,\tilde \epsilon_j)]_{\deg(\epsilon_i,\tilde \epsilon_j)=1} $ extracts from the function $g(\epsilon_i,\tilde \epsilon_j)$ homogeneous terms of degree 1 in $\epsilon_i$ and $\tilde \epsilon_j$ variables around $\tilde\epsilon_j=\infty$, and quantum light dimensions in the heavy-light block are substituted by their classical cousins.   

We note that in the 4-point case the linearized classical block is indeed a linear function in light classical dimensions $\epsilon$ and $\tilde\epsilon$. In the 5-point case, the situation is more intricate as the linearized classical block given as the  perturbative series in the third dimension $\epsilon_3$ also depends on ratios  $(\epsilon_3/\tilde \epsilon_1)^n$ of homogeneity degree 0, cf. \eqref{exf} - \eqref{dddd}. This is why in \eqref{blocks} we extract the homogeneity degree 1 terms.       

Combining the results of the two steps described above we arrive at the following relation
\be
\label{blocks1}
f^{lin}(\epsilon_h, \epsilon_i, \tilde \epsilon_j|z) =  \Big[\log \Big(\prod_i\Big[w^\prime_i(z_i)\Big]^{\epsilon_i}G(\epsilon_h,\epsilon_i, \tilde \epsilon_j|w(z))\Big)\Big]_{\deg(\epsilon, \tilde \epsilon)=1}\;,
\ee 
where $G(\epsilon_h,\epsilon_i, \tilde \epsilon_j|w(z))$ is the global conformal block function, $w_i^\prime(z_i)$ are the Jacobians for the change of  variables of the light external fields,
\be
\label{change0}
(1-z)^\alpha = 1-w\;.
\ee

In the next section, we explicitly consider the 5-point heavy-light conformal block and apply the transition formula \eqref{blocks} to obtain the linearized classical block given by \eqref{exf}. In parallel, we reproduce and discuss the 4-point case.

\section{5-point heavy-light block} 
\label{sec:5hlb}

The heavy fields with dimensions $\Delta_h \equiv \Delta_1=\Delta_2$ are placed in $z_1 = \infty$ and $z_2 = 1$, while light fields with dimensions $\Delta_{3,4,5}$ are in $z_3, z_4$ and $z_5 = 0$. Then, evaluating the light operators in new coordinates 
\be
\label{change}
(1-z_i)^\alpha = 1-w_i\;,\qquad  i =3,4\;,
\ee
we find that the 5-point heavy-light block (modulo a coordinate-independent prefactor) is given by   
\be
\label{generalHL}
\ba{c}
\cV(\epsilon_{h},\Delta_{3,4,5}, \widetilde\Delta_{1,2}|z_3,z_4)\sim  
\Big[w^\prime_3(z_3)\Big]^{\Delta_3}\Big[w^\prime_4(z_4)\Big]^{\Delta_4}\Big[w^\prime_5(0)\Big]^{\Delta_5} G(\Delta_{3,4,5}, \widetilde \Delta_{1,2}|w_3(z_3), w_4(z_4))
\\
\\
\sim (1-w_3)^{\frac{\alpha-1}{\alpha}\Delta_3} (1-w_4)^{\frac{\alpha-1}{\alpha}\Delta_4}  w_3^{\widetilde \Delta_1 -\Delta_3 - \widetilde \Delta_2}  \,w_4^{\widetilde \Delta_2 -\Delta_4 - \Delta_5} F(\Delta_{3,4,5}, \widetilde \Delta_{1,2}|w_3(z_3), w_4(z_4))
\ea
\ee
where  $w^\prime(z)$ is the Jacobian for \eqref{change}, and
\be
G(\Delta_{3,4,5}, \widetilde \Delta_{1,2}|w_3, w_4) = w_3^{\widetilde \Delta_1 -\Delta_3 - \widetilde \Delta_2}  \,w_4^{\widetilde \Delta_2 -\Delta_4 - \Delta_5}F(\Delta_{3,4,5}, \widetilde \Delta_{1,2}|w_3, w_4)\;,
\ee
is the global block contribution to the $5$-point correlator \eqref{series}. In what follows we constraint the conformal dimensions as in \eqref{delta3}. In this case the heavy-light block is given by
\be
\label{generalHLcon}
\ba{c}
\cV(\epsilon_{h},\Delta_{3,4}, \widetilde\Delta_{1}|z_3,z_4)
\sim (1-w_3)^{\frac{\alpha-1}{\alpha}\Delta_3} (1-w_4)^{\frac{\alpha-1}{\alpha}\Delta_4}  w_3^{ -\Delta_3 }  \,w_4^{\widetilde \Delta_1 -2\Delta_4} F(\Delta_{3,4}, \widetilde \Delta_{1,2}|w_3(z_3), w_4(z_4))
\ea
\ee
where $F$ is given by \eqref{222} for $w_{3,4} = w_{3,4}(q_{1,2})$.

\vspace{-3mm}

\paragraph{4-point case.} Setting $\Delta_3 =0$ in \eqref{generalHLcon} we reproduce the  4-point heavy-light  block \cite{Fitzpatrick:2015zha}
\be
\label{fw}
\ba{c}
\cV(\epsilon_{h},\Delta_{4}, \widetilde\Delta_{1}|w) \sim    
(1-w)^{\frac{\alpha-1}{\alpha}\Delta_4} w^{\widetilde\Delta_1-2\Delta_4}\;{}_2F_1(\widetilde\Delta_1,\widetilde\Delta_1, 2 \widetilde\Delta_1 |w)\;.

\ea
\ee

On the other hand, the leading contribution $f^{(0)}$ 
in \eqref{exf} is the $4$-point linearized classical  block \eqref{fcoef}. Using \eqref{lcb} and the change $a = 1-w$, where $a$ is defined in \eqref{fcoef}, the  linearized classical block can be  represented as  
\be
\label{6dec}
 - f^{(0)}(w)= \log\Big[w^{-2\epsilon_4}(1-w)^{\frac{\alpha-1}{\alpha}\epsilon_4}\Big] +\log\Big[ \Big(\frac{\sqrt{1-w}-1}{\sqrt{1-w}+1}\Big)^{\tilde \epsilon_1}\Big]\;.
\ee
In view of \eqref{blocks}, we use \eqref{loghlb} to represent the right-hand side of  \eqref{fw} as
\be
\label{66dec}
\ba{c}
\dps
\log\cV(\epsilon_{h},\epsilon_{4}, \tilde\epsilon_{1}|w) = \log\Big[ w^{-2\epsilon_4} (1-w)^{\frac{\alpha-1}{\alpha}\epsilon_4} \Big]
+\log\Big[w^{\tilde\epsilon_1} \;{}_2F_1(\tilde\epsilon_1,\tilde\epsilon_1, 2 \tilde\epsilon_1 |w) \Big]\;.
\ea
\ee
We see that first terms in \eqref{6dec} and \eqref{66dec} identically coincide while the second ones are apparently different. In particular case of the vacuum block $\tilde \epsilon_1 = 0$ the identification is already achieved. 

When $\tilde \epsilon_1 \neq 0$ we apply  the transition formula \eqref{blocks} and  single out the terms of homogeneity degree 1 in $\tilde \epsilon_1$ (\textit{i.e.} linear in this case). To this end, we expand the logarithms in $w$ and find that the expansion coefficients are generally given by rational functions of $\tilde\epsilon_1$. Indeed, modulo an additive constant we obtain 
\be
\label{log1}
\log\Big[\Big(\frac{\sqrt{1-w}-1}{\sqrt{1-w}+1}\Big)^{\tilde \epsilon_1}\Big] \sim \tilde \epsilon_1  \ln w+\frac{w \tilde \epsilon_1 }{2}+\frac{3 w^2 \tilde \epsilon_1 }{16}+\frac{5 w^3 \tilde \epsilon_1 }{48}+\cO(w^4)\;,
\ee
and  
\be
\label{log2}
\log\Big[w^{\tilde\epsilon_1} \;{}_2F_1(\tilde \epsilon_1,\tilde \epsilon_1, 2 \tilde\epsilon_1 |w) \Big] =\tilde\epsilon_1  \ln w + \frac{\tilde\epsilon_1  w}{2}+\frac{(3 \tilde\epsilon_1^2+2 \tilde\epsilon_1) w^2}{8 (2 \tilde\epsilon_1 +1)}+
\frac{(5 \tilde\epsilon_1^2+4 \tilde \epsilon_1) w^3}{24 (2 \tilde\epsilon_1 +1)}+\cO(w^4)\;.
\ee
According to  \eqref{blocks}, we expand around $\tilde\epsilon_1 = \infty$ in \eqref{log2} and keep  homogeneous terms of order 1 (which in this simple case are linear), and find out that in a given order the resulting series coincide with that  in \eqref{log1}.

It is interesting to note that the above limiting transition can  be readily seen in all orders using the following identity  
\be
\label{hyperexp}
{}_2F_1(\tilde\epsilon_1,\tilde\epsilon_1-\frac{1}{2}, 2 \tilde\epsilon_1 |w) = \Big(\half + \half \sqrt{1-w}\Big)^{1-2 \tilde\epsilon_1}\;,
\ee 
considered in the $\tilde\epsilon_1  = \infty$ limit. 

\vspace{-3mm}

\paragraph{5-point case.}   
Modulo a coordinate-independent prefactor, the logarithm of the heavy-light block \eqref{generalHLcon} in terms of variables $q_1$ and $q_2$ takes the form 
\be
\label{generalHLcon2}
\ba{c}
\log\cV(\epsilon_{h},\Delta_{3,4}, \widetilde\Delta_{1}|q_1,q_2)
=\log\Big[ (1-q_1)^{\frac{\alpha-1}{\alpha}\Delta_3} (1-q_1q_2)^{\frac{\alpha-1}{\alpha}\Delta_4}  q_1^{ -\Delta_3 }  \,
(q_1 q_2)^{\widetilde \Delta_1 -2\Delta_4} F(\Delta_{3,4}, \widetilde \Delta_{1}|q_1, q_2)\Big]\;,
\ea
\ee
where $F(\Delta_{3,4}, \widetilde \Delta_{1}|q_1, q_2)$ is given by \eqref{222}. To compare with the linearized classical block using the transition formula \eqref{blocks} it is convenient to represent \eqref{generalHLcon2} as a power series  in $\epsilon_3$ dimension 
\be
\label{generalHLcon5}
\ba{c}
\dps
\log\cV(\epsilon_{h},\epsilon_{3,4}, \tilde\epsilon_{1}|q_1,q_2)
=\sum_{n=0}^{\infty} g^{(n)}(\epsilon_{h},\epsilon_{4}, \tilde\epsilon_{1}|q_1,q_2) \epsilon_3^n\;,
\ea
\ee
where the lower order coefficients are given by 
\begin{align}
\label{heavy-light-coef}
g^{(0)}= {}& (\tilde\epsilon_1-2\epsilon_4) \log(q_1q_2) +\frac{(\tilde\epsilon_1-2\epsilon_4) q_1 q_2}{2}  + \frac{\tilde\epsilon_1 q_1^2 q_2^2}{4 (1 + 2 \tilde\epsilon_1)} +
\frac{ 3 \tilde\epsilon_1^2 q_1^2 q_2^2}{8 (1 + 2 \tilde\epsilon_1)} - \frac{\epsilon_4 q_1^2 q_2^2}{2} +\nonumber\\
  &
+\frac{ \epsilon_4 q_1 q_2}{\alpha} + \frac{\epsilon_4 q_1^2 q_2^2}{ 2 \alpha} +\cO(q^5)\;,\\
g^{(1)}={}& - \log(q_1)+\frac{q_2-q_1}{2} + \frac{q_2^2-q_1^2}{4 (1 + 2 \tilde\epsilon_1)} - \frac{3 \tilde\epsilon_1 q_1^2}{4 (1 + 2 \tilde\epsilon_1)}  -
 \frac{q_1 q_2}{4}  
  + \frac{\tilde\epsilon_1 q_2^2}{4 (1 + 2 \tilde\epsilon_1)} +\frac{ q_1}{\alpha }+
   \frac{q_1^2}{ 2 \alpha} +\cO(q^3)\;,\\
 g^{(2)}={}&   \frac{q_1^2+q_2^2}{8 (1 + 2 \tilde\epsilon_1)} +\frac{q_1^3+q_2^3}{8 (1 + 2 \tilde\epsilon_1)} - \frac{q_1 q_2(q_1+q_2)}{8 (1 + 2 \tilde\epsilon_1)}  +\cO(q^4)\;,\\
   g^{(3)}={}&   \frac{\tilde\epsilon_1 q_1^4+e q_2^4}{16 (1 + 2 \tilde\epsilon_1)^2 (3 + 2 \tilde\epsilon_1)} - \frac{3 q_1^2 q_2^2}{16 (1 + 2 \tilde\epsilon_1)^2 (3 + 2 \tilde\epsilon_1)} - 
  \frac{\tilde\epsilon_1 q_1^2 q_2^2}{ 8 (1 + 2 \tilde\epsilon_1)^2 (3 + 2 \tilde\epsilon_1)} +\cO(q^5)\;,\\
     g^{(4)}={}& \frac{q_2^4-q_1^4}{64 (1 + 2 \tilde\epsilon_1)^2 (3 + 2 \tilde\epsilon_1)} -\frac{ q_1^5+ q_2^5}{ 32 (1 + 2 \tilde\epsilon_1)^2 (3 + 2 \tilde\epsilon_1)} +
     \frac{q_1q_2(q_1^3+q_2^3)}{ 32 (1 + 2 \tilde\epsilon_1)^2 (3 + 2 \tilde\epsilon_1)}  +\cO(q^6)\;.
\end{align}
This is to be compared to the linearized classical block coefficients \eqref{exf} expanded in $q$-variables 
\begin{align}
\label{linearised-coef}
f^{(0)}={}&(\tilde\epsilon_1-2\epsilon_4) \log(q_1q_2) +\frac{(\tilde\epsilon_1-2\epsilon_4) q_1 q_2}{2} + \frac{3\tilde\epsilon_1 q_1^2 q_2^2}{16}  - \frac{\epsilon_4 q_1^2 q_2^2}{2} +
 \frac{\epsilon_4 q_1 q_2}{\alpha} +\frac{\epsilon_4 q_1^2 q_2^2}{2 \alpha}+\cO(q^5) \;,\\
f^{(1)}={}& - \log(q_1) +\frac{q_2-q_1}{2} - \frac{3 q_1^2}{8} - \frac{
 q_1 q_2}{4} + \frac{q_2^2}{8} + \frac{q_1}{\alpha} + \frac{q_1^2}{2 \alpha}+\cO(q^3) \;,\\
f^{(2)}={}& \frac{1}{4 \tilde\epsilon_1} + \frac{q_1^2}{16 \tilde\epsilon_1} + \frac{q_1^3}{16 \tilde\epsilon_1} - \frac{q_1^2 q_2}{16 \tilde\epsilon_1} + \frac{q_2^2}{16 \tilde\epsilon_1} - 
\frac{q_1 q_2^2}{16 \tilde\epsilon_1} + \frac{q_2^3}{16 \tilde\epsilon_1} +\cO(q^4)  \;,\\
f^{(3)}={}&-\frac{1}{8 \tilde\epsilon_1^2} + \frac{q_1^4}{128 \tilde\epsilon_1^2} 
- \frac{ q_1^2 q_2^2}{64 \tilde\epsilon_1^2} +\frac{ q_2^4}{128 \tilde\epsilon_1^2}
+\cO(q^5)\;,\\
f^{(4)}={}&
\frac{1}{96 \tilde\epsilon_1^3} -\frac{q_1^4}{512 \tilde\epsilon_1^3} -\frac{ q_1^5}{256 \tilde\epsilon_1^3} + \frac{q_1^4 q_2}{ 256 \tilde\epsilon_1^3} - 
\frac{q_2^4}{512 \tilde\epsilon_1^3} + \frac{q_1 q_2^4}{256 \tilde\epsilon_1^3} - \frac{q_2^5}{256 \tilde\epsilon_1^3}+\cO(q^6)\;.
\label{qqqq}
\end{align}
According to  \eqref{blocks}, we expand around $\tilde\epsilon_1 = \infty$ in \eqref{generalHLcon5} keeping  homogeneous terms of order 1 in light dimensions $\epsilon_1$ and $\tilde\epsilon_1$. Up to coordinate-independent terms, the resulting series exactly reproduce those in \eqref{linearised-coef}-\eqref{qqqq}. Note that the 4-point case is obtained here by setting $\epsilon_3 = 0$ that brings us back to $g^{(0)}$ and $f^{(0)}$. Denoting $w = q_1 q_2$ we arrive at \eqref{6dec} and \eqref{66dec} expanded in $w$ ($\tilde\epsilon_1$--dependent terms are given by \eqref{log1} and \eqref{log2}).

\section{Conclusion}
\label{sec:conclusion}

\vspace{3mm}

In this paper we have shown that the two-step method of  Fitzpatrick, Kaplan, and Walters applied to the 5-point global block yields the  5-point linearized classical  block obtained previously in \cite{Alkalaev:2015lca}. To this end, we have explicitly built the 5-point global conformal block using both  the projection technique and the Casimir equations approach.   Our consideration generalizes the FKW method  revealing new features in the case of $n$ points. 

It is worth noting that representing the 5-point global block in as simple a form as possible is still an open problem. For example, in the 4-point case the global block is given by the hypergeometric function which is known to satisfy plenty of useful relations. One of them \eqref{hyperexp} explains the simple form of the 4-point linearized classical block found in \cite{Fitzpatrick:2014vua,Hijano:2015rla}. Among the other  related questions, it would be crucial to formulate the $\epsilon_3$-expansion of the 5-point classical block \eqref{exf} in a closed form. Hopefully, the FKW trick will make it possible to uncover the structures underlying  $n$-point classical conformal blocks with $n$ being a parameter. 

We note that apart from the natural questions like the holographic interpretation of the classical conformal blocks with different portions of heavy and light fields discussed in the introduction there are more speculative but still physically important problems requiring the knowledge of $n$-point conformal blocks in a closed form. For example, having  general $n$-point classical conformal blocks may prove useful in the analysis of the entwinement phenomenon in CFT and its dual interpretation \cite{Balasubramanian:2014sra}. In this case the classical blocks are known to be inappropriate to measure the regions of the angle deficit/BTZ geometry  far enough from the boundary. One possible solution is to consider the so-called "long geodesics" which wrap the defect before returning to the boundary and find corresponding objects in CFT. On the other hand, with $n$ arbitrary we have one more free parameter which can be sent to infinity, thereby producing a particular network of the light particles with total mass comparable to that of the background. Adjusting this parameter to the central charge $c$ may also give other physically interesting phenomena.   

Finally, we may note that besides the AdS/CFT correspondence, the $n$-point analysis in the semiclassical regime is interesting even staying completely inside the conformal field theory. For example, the computation of the classical conformal blocks is sometimes considered as complex Liouville problem in a sense that Liouville action is related to the solution of the accessory parameters problem associated to the uniformization problem related to the real $sl(2,\mathbb{R})$ monodromy group, while in the case of the conformal blocks we are dealing with its complexification $sl(2,\mathbb{C})$. So that the answer to the question what classical system is behind the monodromy problem leading to the classical conformal block   can help to clarify this intriguing connection (see, \textit{e.g.}, \cite{Litvinov:2013sxa}).

\vspace{7mm} 

\noindent \textbf{Acknowledgements.} We thank M. Kalenkov and D. Ponomarev for useful discussions, and J. Kaplan for the correspondence. The work of K.A. was supported by RFBR grant No 14-02-01171. The work of V.B. was performed with the financial support of the Russian Science
Foundation (Grant No.14-50-00150).

\appendix

\section{Global block: details of calculation}
\label{sec:appA}

In what follows we briefly review the computation of the global conformal block of Sec. \bref{sec:proj}. To compute \eqref{formula} we consider the basic matrix element 
\be
\label{AY}
Y(\Delta_1, \Delta_2, \Delta_3| k,m|z) : = \langle \Delta_1| L_1^k \,\psi (z)\, L_{-1}^m | \Delta_3 \rangle\;.
\ee
In particular, setting $m = n=0$ expression \eqref{AY} yields the matrix element associated to the 3-point function, \footnote{The usual normalization is fixed in order to have the leading coefficients of the series expansions of the conformal block equal to one.} 
 $Y(\Delta_1, \Delta_2, \Delta_3| 0,0|z)  = z^{\Delta_1 - \Delta_2 - \Delta_3}$.

The  denominator in \eqref{formula} is computed to be  $\langle  \tilde \Delta_1 |[L_1^m, L_{-1}^m]|  \Delta \rangle = m! (2  \Delta)_m$, where $(a)_k = a(a+1) ... (a+k-1)$ is the Pochhammer symbol, so that the conformal block
can be cast into the form 
\be
\label{AG0}
G(z) = 
\sum_{k,m = 0}^\infty \frac{Y(\Delta_1, \Delta_2, \tilde \Delta_1| 0,k|z_2)\,Y(\tilde \Delta_1, \Delta_3, \tilde \Delta_2| k,m|z_3)Y(\tilde \Delta_2, \Delta_4, \Delta_5| m,0|z_4)}{ (2\tilde \Delta_1)_k (2\tilde \Delta_2)_m\, k! m!}\;,
\ee
where for convenience we relaxed again the value of $z_2$. 

To compute the general matrix element \eqref{AY} we use the $sl(2, \mathbb{C})$ Ward identities 
\be
\label{Award}
[L_k, \phi_{\Delta}(z)] = \big(z^{k+1}\frac{\d}{\d z} + (k+1)\Delta z^k\big)\phi_{\Delta}(z)\;,\qquad  k = -1,0, 1\;,
\ee
to get
\be
\label{AY2}
Y(\Delta_1, \Delta_2, \Delta_3| k,m|z) = \sum_{p = 0}^{\min[k,m]}\,\gamma_p\, \cL_1^{k-p} Y(\Delta_1, \Delta_2, \Delta_3| 0,m-p|z)\;, 
\ee
where operator $\cL_1$  (corresponding  to the Ward identity with $k=1$) is defined as
\be
\cL_1 = z (N_z+2 \Delta_2)\;, \qquad N_z \equiv z\frac{\d}{\d z}\;,
\ee
and the coefficients are given by 
\be
\label{Agamma}
\gamma_p = \frac{k!}{p!(k-p)!} (2\Delta_3 +m-1)^{(p)} m^{(p)}\;,
\ee
where the descending Pochhammer symbol $(a)^{(p)} = a(a-1) ... (a-p+1)$. Note that the first factor in \eqref{Agamma} is just the binomial coefficient.

On the other hand, $Y(\Delta_1, \Delta_2, \Delta_3| 0,s|z)$ in \eqref{AY2} can be computed to be 
\be
Y(\Delta_1, \Delta_2, \Delta_3| 0,s|z) = (-)^s \cL_{-1}^s Y(\Delta_1, \Delta_2, \Delta_3| 0,0|z)\;,
\qquad \quad\cL_{-1} = \frac{\d}{\d z}\;.
\ee 
Substituting the above formula in \eqref{AY2} we obtain 
\be
\label{AY22}
Y(\Delta_1, \Delta_2, \Delta_3| k,m|z) = \sum_{p = 0}^{\min[k,m]}\,(-)^{m-p}\,\gamma_p\, \cL_1^{k-p}\cL_{-1}^{m-p}\, Y(\Delta_1, \Delta_2, \Delta_3| 0,0|z)\;.
\ee

Let us check formula \eqref{AY22}. To this end we set $\Delta_2 = 0$ and the answer in this case has to be  $Y(\Delta_1, \Delta_2, \Delta_3| k,m|z) = \langle \Delta_1| L_1^k  L_{-1}^m | \Delta_3 \rangle = \delta_{\Delta_1 \Delta_3}\delta_{km} \langle \Delta_1| [L_1^k,  L_{-1}^m] | \Delta_3 \rangle  = m! (2\Delta_1)_{m}$. Indeed, if $\Delta_2 = 0$ we find that the matrix element associated with the 3-point function is $\langle \Delta_1 | \mathbb{I} |\Delta_3\rangle  = 1$. It follows that differential operators $\cL_{1, -1}$ acting on the 3-point matrix element are to be of zeroth order which implies $m = p$ and $k=p$ whence  $m=k$. It follows that  $Y(\Delta_1, \Delta_2, \Delta_3| k,m|z) = \gamma_m$, where the coefficient \eqref{Agamma} can be shown to be $\gamma_m = m! (2\Delta_1)_m$. In particular, this consistency check means that inserting the unity operator as the third operator reduces  the original 5-point correlation function to the 4-point correlation function.

Now let us compute degrees of the differential operators $\cL$ in \eqref{AY22}. We find 
\be
\label{ALL}
\ba{l}
\cL_{-1}^s Y(\Delta_1, \Delta_2, \Delta_3| 0,0|z) = (-)^s  (\Delta_3 + \Delta_2 -\Delta_1)_s\, z^{\Delta_1 -\Delta_2 - \Delta_3 - s}\;,
\\
\\
\cL_{1}^p Y(\Delta_1, \Delta_2, \Delta_3| 0,0|z) =  (\Delta_1 + \Delta_2 -\Delta_3)_p \,z^{\Delta_1 -\Delta_2 - \Delta_3 +p}\;,
\\
\\
\cL_1^p \cL_{-1}^s Y(\Delta_1, \Delta_2, \Delta_3| 0,0|z) = 
\\
\\
\hspace{45mm}=(-)^s  (\Delta_3 + \Delta_2 -\Delta_1)_s (\Delta_1 + \Delta_2 -\Delta_3 -s)_p\, z^{\Delta_1 -\Delta_2 - \Delta_3 +p- s}\;,
\ea
\ee
 Using \eqref{ALL} we find that \eqref{AY22} can be cast into the form
\be
\label{AY222}
Y(\Delta_1, \Delta_2, \Delta_3| k,m|z) = 
\tau_{k,m}\, z^{\Delta_1 -\Delta_2 - \Delta_3 +k-m} \;,
\ee
where the coefficient is given by 
\be
\label{Atau}
\tau_{k,m} = \sum_{p = 0}^{\min[k,m]} \frac{k!}{p!(k-p)!} (2\Delta_3 +m-1)^{(p)} m^{(p)}
(\Delta_3+\Delta_2 - \Delta_1)_{m-p}(\Delta_1 + \Delta_2 -\Delta_3+p-m)_{k-p}\;.
\ee
Using \eqref{AY222} we find the final expression for the matrix element \eqref{AG0}:
\be
\label{AG}
\ba{c}
\dps
G(z) = 
 \sum_{k,m = 0}^\infty \frac{(\tilde \Delta_1 + \Delta_2 -\Delta_1)_k\,  \tau_{k,m} \,(\tilde \Delta_2 + \Delta_4 -\Delta_5)_m}{ k! m!\,(2\tilde \Delta_1)_k (2\tilde \Delta_2)_m}\, 
z_2^{\Delta_1 -\Delta_2 - \tilde \Delta_1 - k}\, z_3^{\tilde \Delta_1 -\Delta_3 - \tilde \Delta_2 +k-m}  \,z_4^{\tilde \Delta_2 -\Delta_4 - \Delta_5 +m}\;.
\ea
\ee
To get \eqref{series} we set $z_2=1$.

%

\providecommand{\href}[2]{#2}\begingroup\raggedright
\endgroup

\end{document}